\documentclass[%
 reprint,
showpacs,preprintnumbers,
 amsmath,amssymb,
 aps,
prb,
 longbibliography,
]{revtex4-1}

\usepackage{graphicx}
\usepackage{dcolumn}
\usepackage{bm}
\usepackage{hyperref}
\begin{document}

\preprint{}

\title{
Large Magnetoresistance Effects in $Ln$CoAsO with a \\Ferromagnetic-Antiferromagnetic Transition
}


\author{Hiroto Ohta}
\email[Present address: Institute of Engineering, Division of Advanced Applied Physics, Tokyo University of Agriculture and Technology]{}
\email[]{shioshio@kuchem.kyoto-u.ac.jp}
\thanks{}
\author{Chishiro Michioka}
\author{Kazuyoshi Yoshimura}\email[]{kyhv@kuchem.kyoto-u.ac.jp}

\affiliation{
Department of Chemistry, Graduate School of Science, Kyoto Univeristy, Kyoto 606-8502, Japan
}


\date{\today}

\begin{abstract}
A large magnetoresistance (MR) effect was observed in the layered compounds NdCoAsO and SmCoAsO, in which ferromagnetically ordered itinerant-electrons of Co are sandwiched by localized 4$f$-electrons of $Ln^{3+}$, below ferromagnetic-antiferromagnetic transition (FAFT) temperature $T_\mathrm{N}$ as observed in other FAFT compounds.
In SmCoAsO, the large MR effect is also observed up to the Curie temperature $T_\mathrm{C}$, and it is found to be originating in the presence of another antiferromagnetic phase in the low-magnetic field region of the ferromagnetic phase.
\end{abstract}

\pacs{75.30.Cr, 75.47.-m, 75.50.Cc, 75.60.Ej}

\maketitle


\section{INTRODUCTION}     
\begin{figure}[b]
\begin{center}
\includegraphics[width=6.5cm]{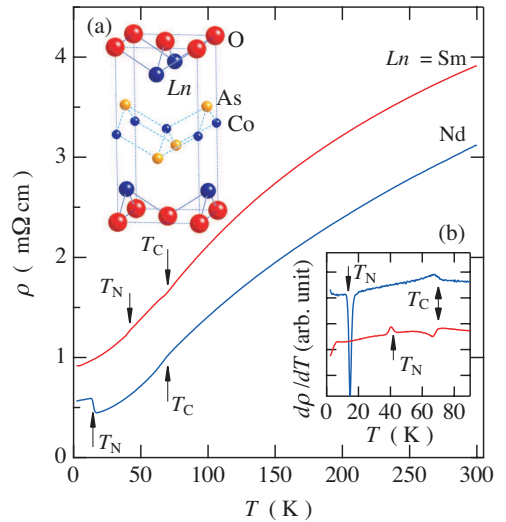} 
\end{center}
\caption{
(color on-line)
$T$ dependence of $\rho$ of NdCoAsO and SmCoAsO at $H$ = 0.
Inset: (a) Crystal structure of $Ln$CoAsO with the space group $P4/nmm$.
(b) $T$ dependence of $d\rho/dT$ of both compounds. 
}
\label{rho_T_0}
\end{figure}

One break-through is always going to induce much influence on the vicinity of it.
The discovery of high-$T_\textrm{c}$ iron pnictides\cite{Kamihara_LaFePO, Kamihara_LaFeAsOF} is such a break-through.
Among the several types of iron based superconductors, 
\cite{Ren_SmFeAsO, Rotter_BaKFe2As2, Wang_LiFeAs, Ogino_Sr4Sc2Fe2P2O6, Hsu_FeSe, Fang_FeTeSePRB} the 1111-type (or ZrCuSiAs type) compounds, including $Ln$FeAs[O$_{1-x}$F$_x$] ($Ln$ equal lanthanoids), have many derivatives with interesting physical properties.
The \mbox{LaCo$Pn$O} ($Pn$ = P and As) compound is one of such derivatives with  itinerant-electron ferromagnetism. \cite{Yanagi_LaCoXO, Sefat_LaCoAsO}
Magnetic property of \mbox{LaCo$Pn$O} can be well understood within the spin-fluctuation theory for the weakly itinerant-electron ferromagnets. \cite{Takahashi_JPSJ1986, Ohta_LaCoAsO}
In the 1111-type compounds, two-dimensional square lattices of transition metal atoms are well separated from each other by LaO layers as seen in the inset (a) of Fig. \ref{rho_T_0}, and thus \mbox{LaCo$Pn$O} are thought to be the ferromagnets with highly 2-dimensional anisotropic magnetic nature. \cite{Ohta_LaCoAsO, Sugawara_LaCoPO}
Substituting La for other lanthanoids, one can introduce magnetic moments of 4$f$-electrons into the in-between layers of Co$Pn$ layers.
The magnetic moments of $Ln^{3+}$ with localized character interact with the ferromagnetically ordered itinerant-electrons of Co through the Ruderman-Kittel-Kasuya-Yoshida (RKKY) interaction. \cite{Krellner_CeCoPO, Ohta_LnCoAsO_mag, Ohta_LaCoAsO_NMR, Marcinkova_NdCoAsO_ND, McGuire_NdCoAsO_ND, Awana_SmCoAsO_mag}
Especially, in the cases of $Ln$ = Nd, Sm, and Gd, the ferromagnetic-antiferromagnetic transitions (FAFT) occur at $T_\mathrm{N}$ = 14 K, 42 K, and about 70 K, respectively, below the Curie temperature ($T_{\textrm{C}}$ $\sim$ 70 K), \cite{Ohta_LnCoAsO_mag, Marcinkova_NdCoAsO_ND, McGuire_NdCoAsO_ND, Awana_SmCoAsO_mag} where $T_\mathrm{N}$ and $T_\mathrm{C}$ are the FAFT temperature and the Curie temperature at $H$ = 0, respectively, in this report.

Recently, McGuire $et$ $al$. reported temperature ($T$) dependence of electric resistivity ($\rho$) of \mbox{NdCoAsO} under the magnetic field ($H$) up to 5 T. \cite{McGuire_NdCoAsO_ND}
In their report, $\rho$ showed an abrupt increase at $T_\mathrm{N}$ with decreasing $T$ as observed in other compounds with FAFT, especially in FeRh and CoMnSi. \cite{Kouvel_JApplPhys1962, Schinkel_JPhysF1974, Zhang_JPhysD2008}
In these FAFT compounds, large magnetoresistance (MR) ratio has been observed below $T_\mathrm{N}$.
The giant MR (GMR) effects are of great importance in the field of device technology because of their application to the large-volume storage and also of interest in the field of fundamental physics since the detailed mechanism of the GMR has still remained unsolved.
Since the roles of localized magnetic moments and itinerant-electron's magnetic moments are well separated in $Ln$CoAsO, it should be less complicate in the mechanism of the MR effect in comparison to above two compounds.
Furthermore, $Ln$CoAsO can be seen as the complete multilayer system with the thickness of one atom size of each magnetic layer.
Therefore, it is of much importance both fundamentally and in application to study the electric resistivity of the newly discovered FAFT compounds, NdCoAsO and SmCoAsO, from the viewpoint of the MR effect.

In this paper, we showed the results of electric resistivity measurements on NdCoAsO and SmCoAsO at various $H$.
We observed the large MR effect below $T_\mathrm{N}$ in both compounds, and also observed it between $T_\mathrm{N}$ and $T_\mathrm{C}$ in SmCoAsO due to the anomalous behavior of $\rho$ at $H$ = 0.
We successfully showed this anomaly in $\rho$ of SmCoAsO originating in the antiferromagnetic phase which is newly found in the low-$H$ region between $T_\mathrm{N}$ and $T_\mathrm{C}$.

\section{EXPERIMENTS}    
For the synthesis of polycrystalline samples of \mbox{$Ln$CoAsO} ($Ln$ = Nd and Sm), we used powders of $Ln$ (purity: 99.9\%), As (99.99\%) and CoO (99.99\%) as starting materials.
The detailed synthesis methods are shown in our previous reports. \cite{Ohta_LnCoAsO_mag, Ohta_SmCoAsO_mag}
$\rho$ of the samples was measured with increasing $T$ under $H$ up to 14 T by a conventional DC four probes method.
For the measurements we used the samples sintered in the shape of a rectangle in size of about 2$\times$2$\times$0.5 mm$^3$.
Magnetization ($M$) of both compounds was measured using the superconducting quantum interference device (SQUID) magnetometer in Research Center for Low Temperature and Materials Sciences, Kyoto University.

\begin{figure}[tb]
\begin{center}
\includegraphics[width=8.5cm]{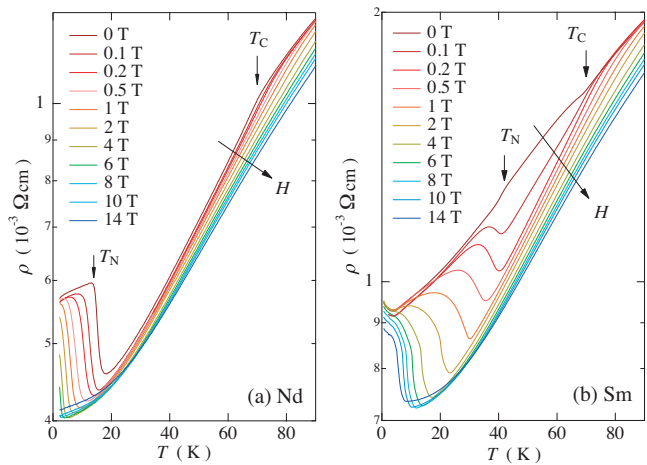} 
\end{center}
\caption{
(color on-line)
$T$ dependence of $\rho$ of (a) NdCoAsO and (b) SmCoAsO at various $H$ from 0 to 14 T.
Arrows indicate $T_\mathrm{C}$ and $T_\mathrm{N}$.
}
\label{rho_T_H}
\end{figure}

\section{Results and Discussion}    
Figure \ref{rho_T_0} shows $T$ dependence of $\rho$ at $H$ = 0.
Both compounds show metallic conduction in all temperature regions, being consistent to LaCo$Pn$O. \cite{Yanagi_LaCoXO}
In NdCoAsO, there have been seen a slight decrease at $T_\mathrm{C}$ and an abrupt increase, or a ``jump", at $T_\mathrm{N}$ with decreasing $T$ in $\rho$ as reported by McGuire $et$ $al$. \cite{McGuire_NdCoAsO_ND}
This behavior can also be observed in $d\rho/dT$ as shown in the inset (b) of Fig. \ref{rho_T_0} more clearly.
The same behavior was reported in FeRh and CoMnSi, \cite{Kouvel_JApplPhys1962, Schinkel_JPhysF1974, Zhang_JPhysD2008} indicating these characters to be common nature of the FAFT compounds.
On the other hand, the opposite behavior, i.e., a slight increase at $T_\mathrm{C}$ and a slight decrease at $T_\mathrm{N}$ was observed in $\rho$ of SmCoAsO.
This ``bump"-like behavior of $\rho$ indicates the different electronic state is realizing between $T_\mathrm{N}$ and $T_\mathrm{C}$ in SmCoAsO.

$T$ dependence of $\rho$ at various $H$ from 0 to 14 T is shown in Fig. \ref{rho_T_H}.
In the case of $Ln$ = Nd, the ``jump" of $\rho$ observed at $T_\mathrm{N}$ shifts to the low-$T$ direction.
This is consistent with the previous report\cite{McGuire_NdCoAsO_ND} and with our report of magnetization. \cite{Ohta_LnCoAsO_mag}
Around $T_\mathrm{C}$, $\rho$ is reduced by $H$.
Such a reduction of $\rho$ can be understood as that ferromagnetic fluctuations are reduced by $H$.
In the case of $Ln$ = Sm, except for $H$ = 0, the behavior of $\rho$ under $H$ was similar to that of NdCoAsO: 
the ``jump" of $\rho$ was observed and its temperature decreased with increasing $H$, and $\rho$ was reduced by $H$ around $T_\mathrm{C}$.
Due to the ``bump" of $\rho$ between $T_\mathrm{N}$ and $T_\mathrm{C}$, $\rho$ of SmCoAsO shows strong $H$ dependence in the low-$H$ region of this $T$ region.

\begin{figure}[tb]
\begin{center}
\includegraphics[width=7cm]{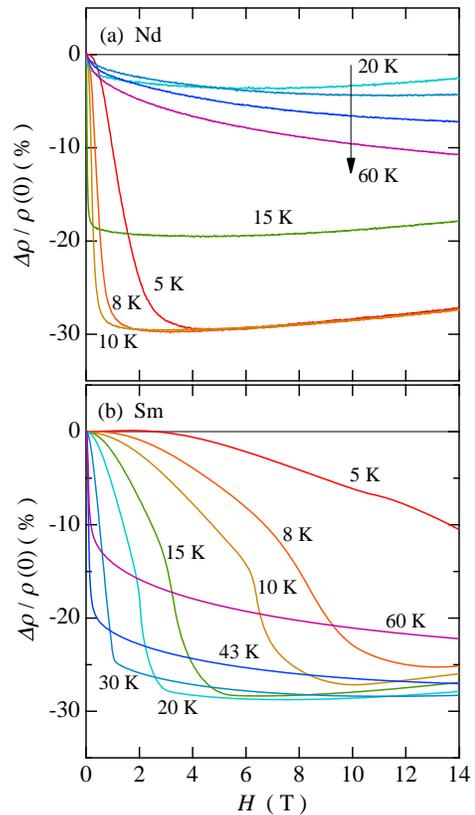} 
\end{center}
\caption{
(color on-line)
$H$ dependence of $\Delta\rho/\rho(0)$ of (a) NdCoAsO and (b) SmCoAsO at various $T$.
}
\label{delta_rho}
\end{figure}

\begin{figure}[tb]
\begin{center}
\includegraphics[width=9.5cm]{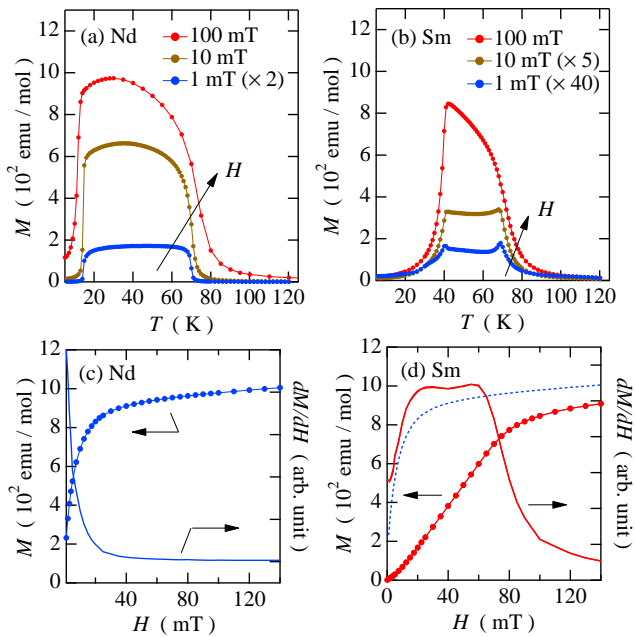} 
\end{center}
\caption{
(color on-line)
$T$ dependence of $M$ of (a) NdCoAsO and (b) SmCoAsO at $H$ = 1, 10 and 100 mT.
The data at 1 and 10 mT of SmCoAsO and those at 1 mT of NdCoAsO are magnified for a good view. 
$H$ dependence of $M$ and $dM/dT$ of (c) NdCoAsO at 40 K and (d) SmCoAsO at 43 K.
Dotted line in the panel (d) shows the $M$ of NdCoAsO.
}
\label{MT}
\end{figure}

To see the detailed change of $\rho$ against $H$, we measured $H$ dependence of $\rho$ up to 14 T at various $T$.
Figure \ref{delta_rho} shows $H$ dependence of MR ratio ($\Delta\rho/\rho(0)$) at various $T$ from 5 K to 60 K, where $\Delta\rho = \rho(H) - \rho(0)$ and $\rho(H) = \rho$ at $H$.
In the case of $Ln$ = Nd, large values of MR ratio ($\sim$ -30 \%) were observed below $T_\mathrm{N}$, and just above $T_\mathrm{N}$ the value of about -20 \% was achieved.
In addition, such large values were easily achieved below 1 T in this $T$ region.
On the other hand, above 20 K, where there is not influence of FAFT but that of ferromagnetic-fluctuations, the values of MR ratio were not so large and showed weak dependence on $H$.
Thus it needs high $H$ to obtain the high values of MR ratio in this $T$ region.
In the case of $Ln$ = Sm, somewhat different behavior was observed as seen in Fig. \ref{delta_rho} (b).
Below $T_\mathrm{N}$, large values of MR ratio ($\sim$ -28 \%) were obtained as in the case of $Ln$ = Nd although rather high $H$ was needed compared with the Nd case.
Above 43 K, the MR ratio still remained in a high value, and the value over -10 \% can be achieved below 1 T in this $T$ region.
This difference from the Nd case is clearly originating in the ``bump" observed in $T$ dependence of $\rho$(0) between $T_\mathrm{N}$ and $T_\mathrm{C}$, since $H$ dependence of MR ratio in this $T$ region showed the same behavior in the case of Nd above $H$ = 1 T.
Due to this anomalous ``bump" of $\rho$ SmCoAsO comes to show large MR ratio in much wider $T$ region compared with NdCoAsO.

\begin{figure}[tb]
\begin{center}
\includegraphics[width=7cm]{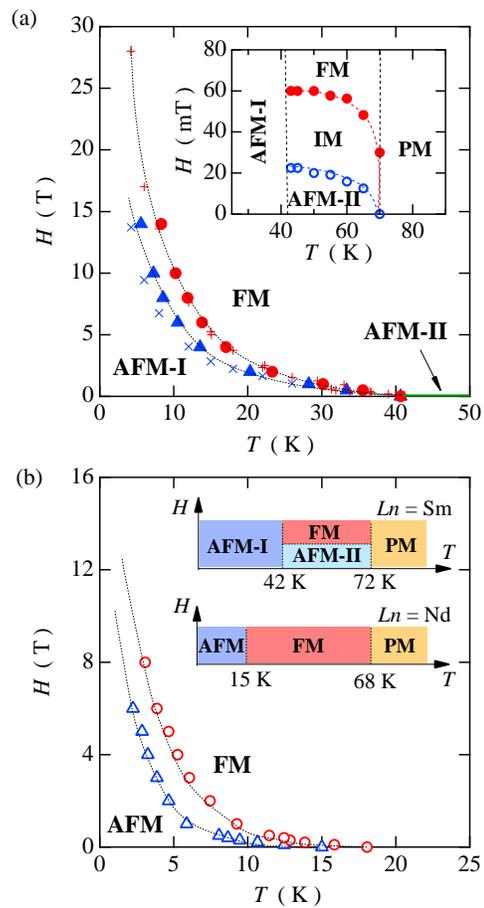} 
\end{center}
\caption{
(color on-line)
(a) Magnetic phase diagram of SmCoAsO.
Circles and triangles show the temperatures where $d\rho/dT$ takes zero and the minimum, respectively.
Small ``+" and ``$\times$" symbols stand for $H$ where $dM/dH$ shows anomaly. \cite{Ohta_SmCoAsO_mag}
Dotted lines are guides for the eyes.
AFM-I and FM denote the antiferromagnetically and ferromagnetically ordered phases.
AFM-II denotes the newly discovered antiferromagnetic phase at the low-$H$ region.
Inset: the low-$H$ region of phase diagram.
Open and closed circles show the values estimated by the same manner in the case of SmCoAsO.
IM and PM denote intermediate state during metamagnetic transition and paramagnetic phase.
(b) Magnetic phase diagram of NdCoAsO.
Circles and triangles show the temperatures where $d\rho/dT$ takes zero and the minimum, respectively.
AFM denotes the antiferromagnetically ordered phase.
Inset: schematic phase diagrams of both compounds in the low-$H$ region.
}
\label{PD}
\end{figure}

To clarify what happens in the low-$H$ region between $T_\mathrm{N}$ and $T_\mathrm{C}$ in SmCoAsO, we measured magnetization $M$ in the low-$H$ region presicely.
Figures \ref{MT} (a) and \ref{MT} (b) show $T$ dependence of $M$ of NdCoAsO and SmCoAsO, respectively, at $H$ = 1, 10, and 100 mT.
Since $M$ of SmCoAsO at $H$ = 1 and 10 mT and that of NdCoAsO at 1 mT are too small to show in the same scale with other data, we magnified these data as indicated in the figures. 
In the case of $Ln$ = Nd, a ferromagnetic feature, i.e., a convex curvature with abrupt changes at $T_\mathrm{N}$ and $T_\mathrm{C}$, was observed down to 1 mT between $T_\mathrm{N}$ and $T_\mathrm{C}$, while in the case of $Ln$ = Sm, an obvious anomaly, i.e., a concave curvature, was observed in the corresponding $T$ region below 10 mT.
This indicates that ferromagnetically ordered magnetic moments at $T_\mathrm{C}$ are immediately forced to change their arrangement of moments to antiferromagnetic one.
The same anomalous behavior of $M$ in the low-$H$ region was also reported in Ref. [19], showing that this anomaly comes from the intrinsic nature of SmCoAsO, not from impurities.
Fig. \ref{MT} (c) and \ref{MT} (d) show isothermal magnetic curves at 40 K for the Nd case and at 43 K for the Sm case, where both temperatures correspond to the ``ferromagnetic" phase.
 In contrast to the case with $Ln$ = Nd, in which $M$ shows ferromagnetic $H$-dependence in all the $H$-region, in the case of $Ln$ = Sm, a concave curvature, which reminds us of a metamagnetic transition, was observed up to 80 mT, showing the existence of an antiferromagnetic phase.
This antiferromagnetic phase is easy to break under weak $H$, and above 80 mT the itinerant-electronic system of SmCoAsO behaves as ferromagnetic one in this $T$-region.

In Fig. \ref{PD} (a), we summarized the results of measurements as the magnetic phase diagram of SmCoAsO.
Closed circles and triangles respectively show the temperatures where $d\rho/dT$ comes to zero and the minimum at various $H$.
Small ``+" and ``$\times$" symbols show the values of $H$ at which $dM/dH$ shows a discontinuity. \cite{Ohta_SmCoAsO_mag}
Good consistency between the estimated values from the results of electric resistivity and magnetization shows that the ``jump" of $\rho$ corresponds to the FAFT, which was confirmed through magnetization measurements.
The obtained phase diagram resembles that of CoMnSi, \cite{Sandeman_PRB2006} indicating the mechanism of FAFT being similar to each other.
Inset of Fig. \ref{PD} (a) shows the low-$H$ region of the phase diagram of SmCoAsO.
Open and closed circles show the values of $H$ corresponding to the upper and lower peaks of $dM/dH$, respectively.
It is interesting that both FM and AFM-II phases are on the border with AFM-I phase at $T_\mathrm{N}$ = 42 K.
The magnetic phase diagram of NdCoAsO is quite similar to that of SmCoAsO as shown in Fig. \ref{PD} (b), except for the absence of AFM-II phase.
The phase diagrams of both compounds are schematically shown in the inset of Fig. \ref{PD} (b) for a good understanding of the similarity and difference between two compounds.
It should be noted that another magnetically ordered phase is reported in the low-$T$ region of the AFM phase. \cite{Marcinkova_NdCoAsO_ND, McGuire_NdCoAsO_ND, Awana_SmCoAsO_mag}
However, we omitted it from Fig. \ref{PD} for simplicity.

As observed in Fig. \ref{rho_T_H} (b), $\rho$ showed an increase with decreasing $T$ below 5 K in SmCoAsO.
Recently, another antiferromagnetic phase was supposed to exist below about 5 K. \cite{Awana_SmCoAsO_mag}
Therefore, the increase of $\rho$ below 5 K may be closely related to the low-$T$ antiferromagnetic phase.
It is an interesting and challenging issue although we did not have much information of the low-$T$ antiferromagnetic phase at this stage.

From the viewpoint of crystal structure, $Ln$CoAsO can be seen as a ``natural" magnetic thin-film system: each ``film" precisely has a thickness of one atom size, which is in a sequence of Co-$Ln$-$Ln$-Co-$\cdots$, where magnetic moments lie within each plane.
This is the situation similar to the model discussed in the review for artificial multilayer systems by Camley and Stamp, \cite{Theory_MR} and the behavior of $\rho$ in $Ln$CoAsO seems to be explained by the spin-dependent scattering, i.e., the ``jump" of $\rho$ with decreasing $T$ is explained to be caused by the change of arrangement of magnetic spins from ferromagnetic to antiferromagnetic.
If this scenario is correct, the component of $\rho$ responsible to FAFT can be ascribed to the component along the $c$-axis.
In $Ln$CoAsO, however, electric conductivity along the $c$-axis is thought to be bad because of two-dimensional band structures, and thus resistivity along the $c$-axis should be masked by that in the $ab$-plane in our measurement condition.
For further detailed study to clarify the mechanism of large MR effects in $Ln$CoAsO, single crystalline samples are needed.
Fortunately, the knowledge obtained through the study of iron-pnictide superconductors must be available for the synthesis of the single crystal. \cite{Ishikado_1111single}

\section{CONCLUSION}
In summary, we measured temperature dependence of electric resistivity of NdCoAsO and SmCoAsO at various magnetic field up to 14 T.
In addition to the abrupt increase in electric resistivity at the FAFT temperature, SmCoAsO shows an anomalous behavior, which is clarified to be originating in the newly antiferromagnetic phase in the low magnetic field region.
Due to this antiferromagnetic phase, large magnetoresistance ratio was realized even   above the FAFT temperature up to the Curie temperature in SmCoAsO.

\begin{acknowledgments}
This work is supported by Grants-in-Aid for Scientific Research from the Japan Society for Promotion of Science (Grants No. 19350030 and No. 22350029).
\end{acknowledgments}

\bibliography{LnCoAsO_R}

\end{document}